\documentclass{desyproc}

\begin{document}
\title{New PVLAS model independent limit for the axion coupling to $\gamma\gamma$ for axion masses above 1~meV}

\author{{\slshape F. Della Valle$^1$, A. Ejlli$^{2}$, U. Gastaldi$^{3}$, G. Messineo$^{2}$, E. Milotti$^{1}$, R. Pengo$^{4}$, L. Piemontese$^{2}$, G. Ruoso$^{4}$, G. Zavattini$^{2}$ }\\[1ex]
$^1$ INFN, Sezione di Trieste and Dipartimento di Fisica, Universit\`{a} di Trieste, Italy\\
$^2$ INFN, Sezione di Ferrara and Dipartimento di Fisica, Universit\`{a} di Ferrara, Italy\\
$^3$ INFN, Sezione di Ferrara, Ferrara, Italy\\
$^4$ INFN, Laboratori Nazionali di Legnaro (PD), Italy}

\contribID{familyname\_firstname}


\maketitle

\begin{abstract}
During 2014 the PVLAS experiment has started data taking with a new apparatus installed at the INFN Section of Ferrara, Italy. The main target of the experiment is the observation of magnetic birefringence of vacuum. 
According to QED, the ellipticity generated by the magnetic birefringence of vacuum in the experimental apparatus is expected to be $\psi^{\rm(QED)} \approx 5\times10^{-11}$.  No ellipticity signal is present so far with a noise floor $\psi^{\rm(noise)} \approx 2.5\times10^{-9}$ after 210 hours of data taking. The resulting ellipticity limit provides the best model independent upper limit on the coupling of axions to $\gamma\gamma$ for axion masses above $10^{-3}$~eV.
\end{abstract}

\section{Introduction}
Several experimental efforts have been set up with the main goal of observing magnetic birefringence of vacuum~\cite{Cameron93,Zavattini08,NJP13}. This effect is expected to generate ellipticity on a linearly polarized  light beam which propagates in vacuum in the presence of a magnetic field ${\bf B}$ orthogonal to the direction of the light beam.  Magnetic birefringence of vacuum is a manifestation of nonlinear electrodynamic effects~\cite{QED} due to vacuum fluctuations and predicted before the full formulation of QED. It results from the interaction of incoming laser photons with virtual photons of the magnetic field. 
Magnetic birefringence of vacuum is closely related to elastic light-by-light interaction~\cite{Bernard}; neither effect has yet been observed. Detection of the magnetic birefringence of vacuum is of major 
importance because it would represent the first direct observation of the interactions between gauge bosons present both in the initial and the final states. 


Spin zero light particles which couple to two photons may also contribute to birefringence~\cite{axion}: observation of a magnetic birefringence different from the value predicted by QED could indicate the existence of light scalar or pseudoscalar particles [axions~\cite{axion2} or axion-like particles (ALPs)] of paramount importance for cosmology and for QCD. A limit on the ellipticity generated in an apparatus like the PVLAS ellipsometer gives a model independent indication on the axion mass $m_a$ and the coupling constant $g_a$ to two photons. Much better limits reported by CAST~\cite{CAST} depend, instead, on the assumed density of spin zero particles that traverse the apparatus.

\section{Apparatus and Method}
\begin{figure}[htb]
\centering\includegraphics[width=0.8\linewidth]{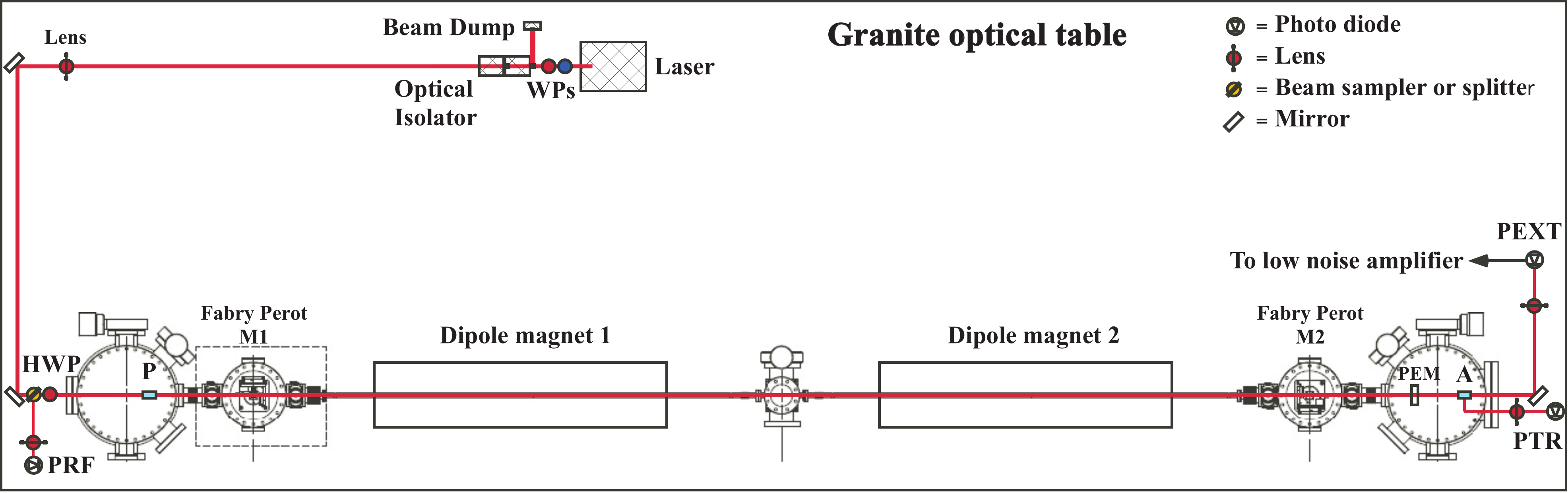}\\
\centering\includegraphics[width=0.8\linewidth]{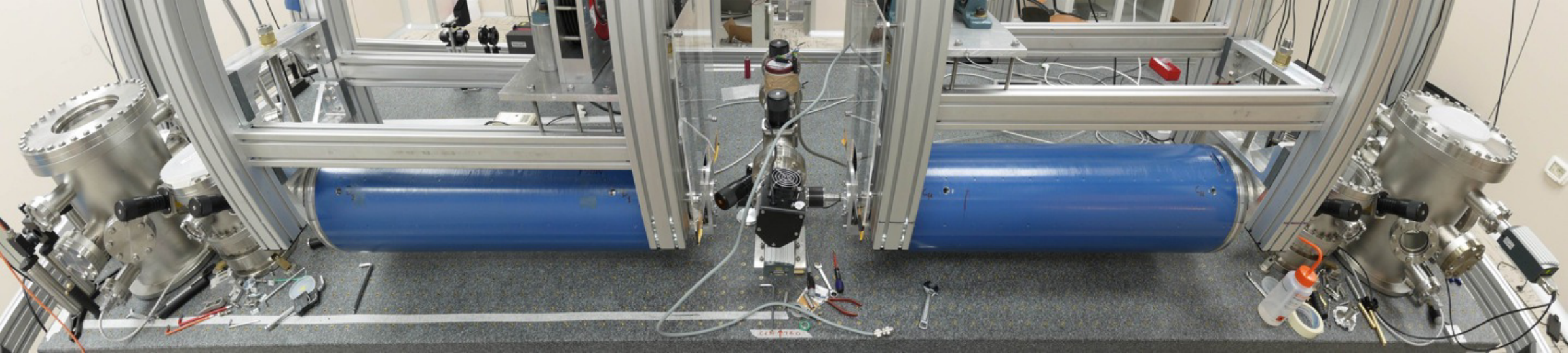}
\caption{Upper panel: scheme of the apparatus. 
Lower panel: wide-angle picture of the PVLAS apparatus. The two blue cylinders are the dipole permanent magnets.}
\label{apparato}
\end{figure}

A new set up is installed in Ferrara, Italy, featuring a tabletop ellipsometer with two identical permanent dipole magnets each with a maximum field of 2.5 T over a 0.8 m length rotating up to 10~Hz. Figure~\ref{apparato}, top panel, gives a schematic top view of the apparatus, whereas Fig.~\ref{apparato}, lower panel, shows a photograph of the apparatus. 
The principle of the experiment follows the measurement scheme proposed in 1979 by Iacopini and Zavattini \cite{iacopini}. Linearly polarized light is injected in the ellipsometer, which is installed in a UHV enclosure. The ellipsometer consists of an entrance polarizer P and an analyzer A set to maximum extinction. Between P and A are installed the entrance mirror M1 and the exit mirror M2 of a Fabry-Perot cavity FP with ultra-high finesse $\cal F$ \cite{recordfinesse}.
The laser source is frequency locked to the cavity~\cite{PDH}.
The light stored between the two FP mirrors travels through the magnetic field region making typically $N = \frac{2{\cal F}}{\pi} = 4.3 \times 10^5$ reflections (corresponding to a path of about 800~km). 
When the fields of the two magnets are parallel, the ellipticity $\psi$ induced on the linearly polarized light beam is 
\[\psi = N\psi_{\rm single} = N\frac{\pi \Delta n_{u}^{\rm (vac)}\int_{0}^{L}{B^2 dl}}{\lambda}\sin{2\vartheta} \approx 5 \times 10^{-11}\:\sin{2\vartheta}\]
where $\lambda=1064$~nm is the laser wavelength, $L=1.6$~m the length of the path through the two magnets, $\vartheta$ is the angle between the light polarization vector and the magnetic field vector and $\Delta n_u^{\rm (vac)} = 3.97\times 10^{-24}$~T$^{-2}$ is the unitary birefringence of vacuum predicted by QED.
To measure such a small quantity,
a heterodyne technique is used: a photoelastic modulator PEM introduces a time dependent carrier ellipticity with amplitude $\eta \approx 10^{-3}$ at a frequency $\nu_{\rm PEM} \approx 50$~kHz and the magnets are set in rotation at the same frequency $\nu_{\rm mag} \approx 3$~Hz. The ellipticity generated by the magnets is modulated at twice the magnet rotation frequency. In the frequency spectrum of the signals detected by the photodiode PEXT the beating of the magnet and PEM ellipticities generate signals with amplitudes proportional to $2\eta\psi \approx 10^{-13}$ at the frequencies $\nu_{\rm PEM} \pm 2\nu_{\rm mag}$ (see ref.~\cite{NJP13}). 


\section{Results}
\begin{figure}[htb]
\centering\includegraphics[width=0.5\linewidth]{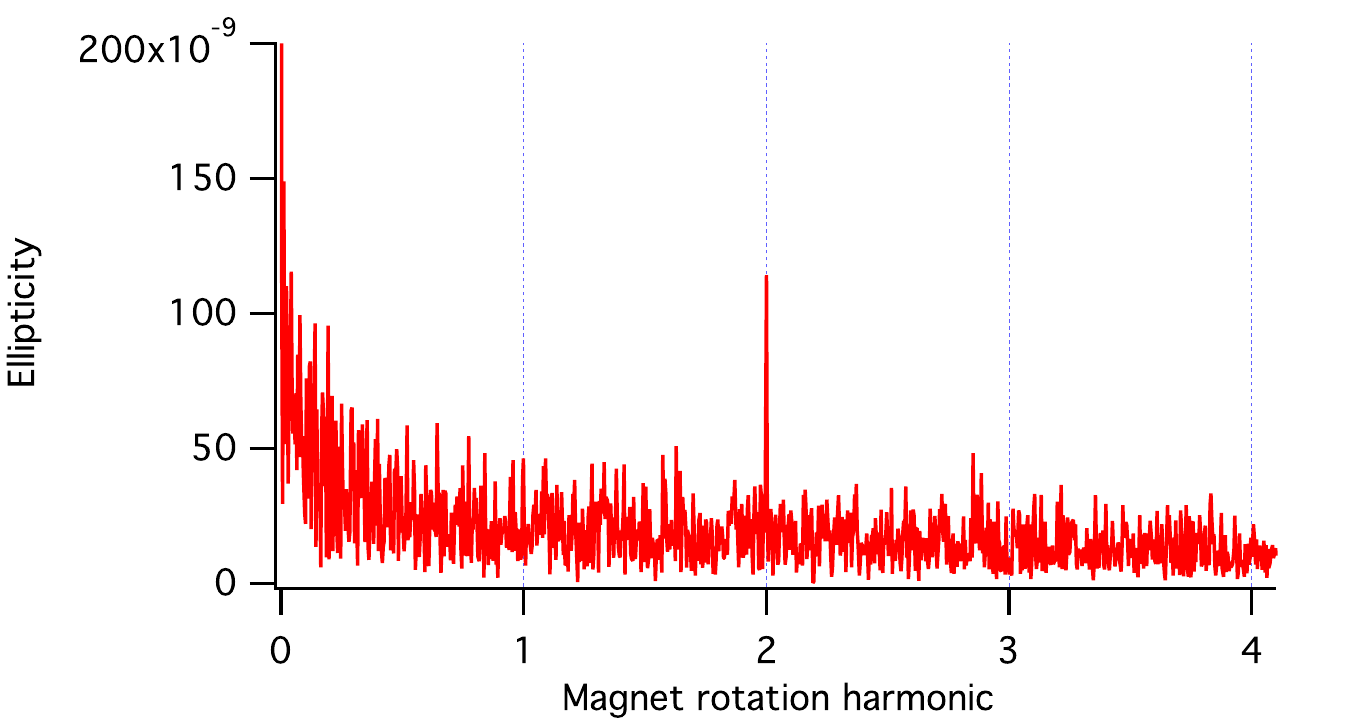}\includegraphics[width=0.5\linewidth]{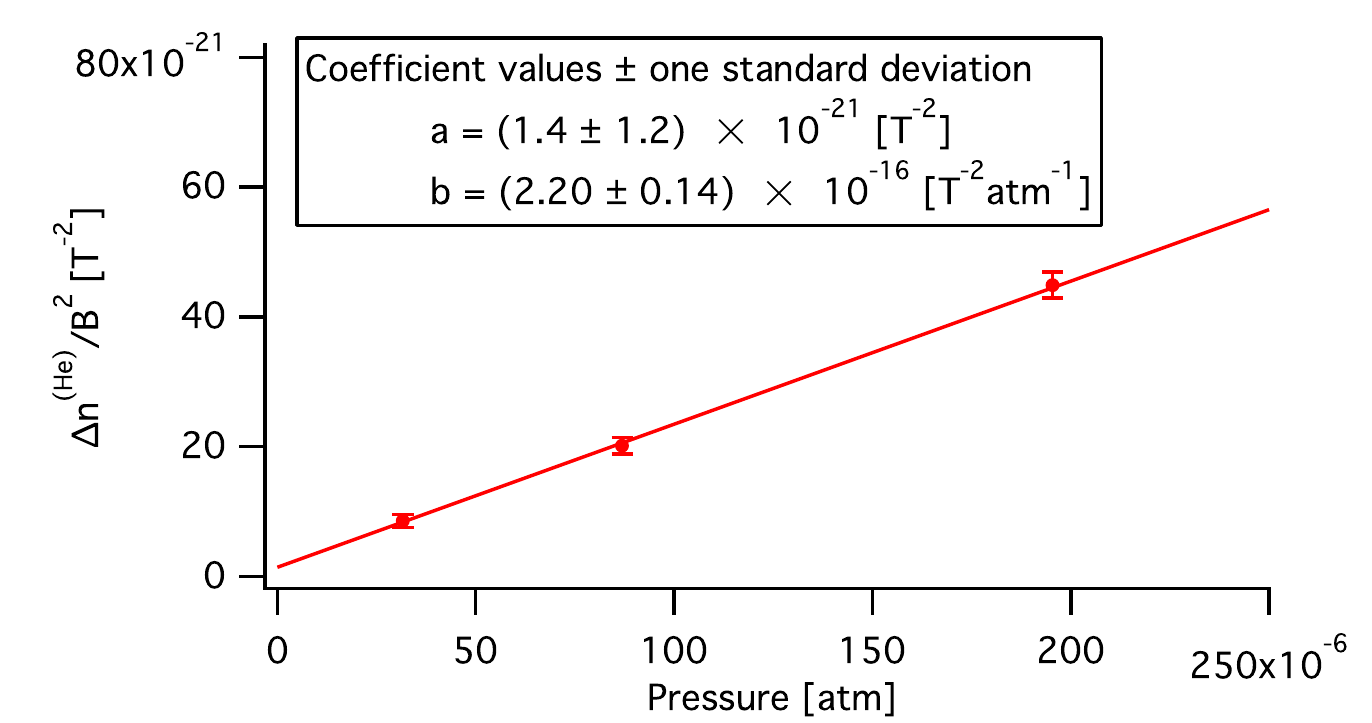}
\caption{Left: Fourier spectrum of the measured ellipticity $\psi(t)$ with 32~$\mu$bar pressure of He after demodulation at $\Omega_{\rm PEM}$. The integration time was T = 5.7 hours.
The vacuum magnetic birefringence predicted by QED is equivalent to a He pressure of $\sim 20$~nbar. Right: measured ${\Delta n^{\rm (He)}}/{B^2}$ as a function of pressure ${P}$. The error bars correspond to a 1$\sigma$ statistical error. The data are fitted with a linear function $a+b{P}$.}
\label{He32}
\end{figure}

The calibration of the apparatus has been done by measuring the Cotton-Mouton effect  of helium gas at several pressures and controlling the consistency of the results with the values present in the literature.  
The left panel of Fig.~\ref{He32} shows the Fourier spectrum of PEXT signals demodulated around $\nu_{\rm PEM}$ for a 5.7 hours run with 32 $\mu$bar He in the UHV enclosure and the magnets rotating at $\nu_{\rm mag} = 3$~Hz.  The spectrum features a clear peak at  $2\nu_{\rm mag}$ corresponding to a Cotton-Mouton ellipticity $\psi{\rm (He\;@\;32 \mu bar)} =1.13\times 10^{-7}$. The baseline ellipticity noise around 6 Hz is of the order of  $1.5\times 10^{-8}$. The right panel of Fig.~\ref{He32} shows the ellipticity signal generated by He at three different pressures.

With the ellipsometer in vacuum 210 hours of data have been analyzed, taking into account amplitude and phase for the signal at $2\nu_{\rm mag}$ for each run \cite{submitted}. No peak is present in the resulting  spectrum. The baseline noise level around  $2\nu_{\rm mag}$ is $\psi_{\rm noise} \approx 2.5\times 10^{-9}$.  
Although this represents a major improvement compared to previous measurements, in view of the expected sugnal of the magnetic birefringence of vacuum $\psi \approx 5\times 10^{-11}$, we have still to recover a missing factor $\approx 50$. This will be done identifying and reducing the sources of noise, which is presently far above the shot-noise limit.

\begin{figure}[h]
\centering\includegraphics[width=0.65\linewidth]{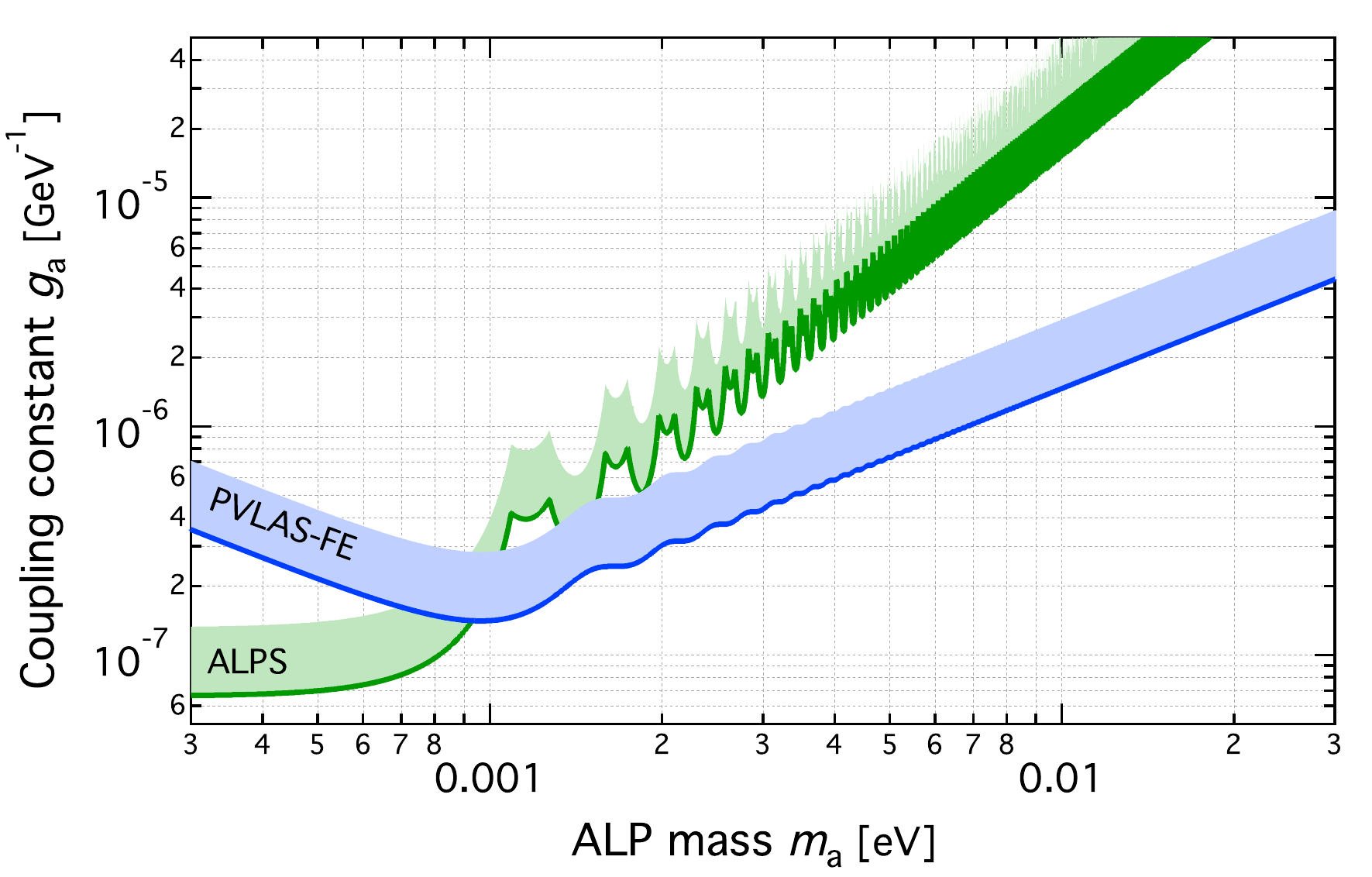}
\caption{Exclusion plot at 95 \% c.l. for axion-like particles for model independent experiments. In green, limits from the ALPS collaboration \cite{ALPs}; in blue the new bounds presented here.}
\label{alp}
\end{figure}
The ellipticity induced by low mass axions can be expressed as \cite{Cameron93}:
\[\psi_{\rm axion} = \frac{N\pi L}{\lambda}\,\frac{g_a^2B^2}{2m_a^2}\left(1-\frac{\sin{2x}}{2x}\right)\]
where $m_a$ is the axion mass, $L$ the magnetic field length, $x = \frac{L m_a^2}{4\omega}$  and $\omega$ is the energy of the laser photons.
The measured ellipticity noise $\psi_{\rm noise}\approx 2.5 \times10^{-9}$ gives an upper limit  for $\psi_{\rm axion}$, represented as an exclusion plot for the coupling constant $g_a$ as a function of the axion mass $m_a$.  Figure~\ref{alp} updates the limits given by model independent laser experiments. 
One may notice that the limits derived by PVLAS-Fe are the most restrictive in the $m_a$ mass region above $10^{-3}$~eV.

 



\begin{footnotesize}

\end{footnotesize}
\end{document}